\documentclass[sigplan,10pt]{acmart}
\newcommand{\revision}[1]{#1}
\newcommand{\shepherd}[1]{#1}

\usepackage{soul}
\soulregister\cite7
\soulregister\ref7

\usepackage{tikz}
\definecolor{lightmintbg}{rgb}{1,1,1}
\newcommand*\circleint[1]{\tikz[baseline=(char.base)]{
            \node[shape=circle,draw,inner sep=2pt, fill=lightmintbg] (char) {\footnotesize #1};}}

\usepackage{subcaption}
\usepackage[]{hyperref}
\usepackage{algorithmic}
\usepackage{graphicx}
\usepackage{textcomp}
\usepackage{fancyhdr}
\usepackage[x11names]{xcolor}
\usepackage{listings}
\usepackage{tcolorbox}
\usepackage{multirow}
\usepackage{booktabs}
\usepackage{enumitem}
\usepackage{makecell}
\usepackage{pifont}
\usepackage{flushend}
\usepackage{changepage}
\usepackage{balance}
\usepackage{float}

\AtBeginDocument{%
  }

\begin{document}

\ccsdesc[300]{Computer systems organization~Cloud computing}
\ccsdesc[300]{Software and its engineering~Virtual machines}
\ccsdesc[500]{Software and its engineering~Virtual memory}
\ccsdesc[500]{Software and its engineering~Allocation / deallocation strategies}

\keywords{Cloud Computing, Serverless, FaaS, Virtualization, Memory Elasticity, Ballooning, Memory Hotunplug, Partitioning, Allocation}
\acmSubmissionID{589}

\copyrightyear{2026}
\acmYear{2026}
\setcopyright{cc}
\setcctype{by}
\acmConference[EUROSYS '26]{21st European Conference on Computer Systems}{April 27--30, 2026}{Edinburgh, Scotland Uk}
\acmBooktitle{21st European Conference on Computer Systems (EUROSYS '26), April 27--30, 2026, Edinburgh, Scotland Uk}
\acmPrice{}
\acmDOI{10.1145/3767295.3769357}
\acmISBN{979-8-4007-2212-7/26/04}

\title{Squeezy: Rapid VM Memory Reclamation for Serverless Functions}

\author{Orestis Lagkas Nikolos}
\email{olagkas@cslab.ece.ntua.gr}
\affiliation{%
  \institution{National Technical University Of Athens}
  \city{Athens}
  \country{Greece}
}

\author{Chloe Alverti}
\email{xalverti@illinois.edu}
\affiliation{%
  \institution{University of Illinois Urbana-Champaign}
  \city{Illinois}
  \country{USA}
}

\author{Stratos Psomadakis}
\email{psomas@cslab.ece.ntua.gr}
\affiliation{%
  \institution{National Technical University Of Athens}
  \city{Athens}
  \country{Greece}
}

\author{Georgios Goumas}
\email{goumas@cslab.ece.ntua.gr}
\affiliation{%
  \institution{National Technical University Of Athens}
  \city{Athens}
  \country{Greece}
}

\author{Nectarios Koziris}
\email{nkoziris@cslab.ece.ntua.gr}
\affiliation{%
  \institution{National Technical University Of Athens}
  \city{Athens}
  \country{Greece}
}

\renewcommand{\shortauthors}{Lagkas Nikolos, et al.}

\begin{abstract}
Resource elasticity is one of the key defining characteristics of the Function-as-a-Service (FaaS) serverless computing paradigm.
While compute resources assigned to VM-sandboxed functions can be seamlessly adjusted on the fly, memory elasticity remains challenging.
Hot(un)plugging memory resources suffers from
long reclamation latencies and occupies valuable CPU resources.
We identify the obliviousness of the OS memory manager to the  
hotplugged memory as the key issue hindering hot-unplug performance, and design Squeezy, a novel approach for fast and efficient VM memory hot(un)plug, targeting VM-sandboxed serverless functions.
Our key insight is that by segregating  
hotplugged memory regions from regular VM memory, we are able to bound the 
lifetime of allocations within these regions thus enabling their fast and efficient reclamation. 
We implement Squeezy in Linux v6.6 as an extension to the OS memory manager. 
Our evaluation reveals that Squeezy
is an order-of-magnitude faster than state-of-the-art, keeping tail latency bounded, when
reclaiming VM memory, achieving sub-second reclamation
of multiple GiBs of memory while serving realistic FaaS load.
\end{abstract}

\maketitle

\vspace*{-5.0mm}
\section{Introduction}

One of the defining features of Function-as-a-Service\cite{awslambda,azure,googleserverless,alibaba,huawei,cloudflare} (FaaS) serverless computing paradigm is the pay-as-you-go model, 
in which providers execute functions on-demand and 
charge users only for the resources consumed during function execution.  
This critical feature requires up- and down-scaling compute and memory 
resources dynamically and transparently,   
based on the incoming load.
Providers typically\textbf{} scale the resources
by increasing or decreasing the number 
of function instances that serve requests.
Ideally, this would be combined with the
dynamic and synchronous
allocation and release of their 
resources, i.e., memory and CPU. 
Unfortunately, such resource agility  
is difficult to achieve while guaranteeing performance objectives. 

The performance tax of resource elasticity in FaaS is
highly dependent on the isolation model used.
Functions are typically sandboxed within containers~\cite{docker}
deployed on  
virtual machines (VMs)~\cite{firecracker}. 
FaaS providers
may choose to a) deploy each function instance in a new
dedicated microVM, i.e., the \emph{single-container-per-VM} model (1:1)
adopted by AWS Lambda~\cite{awslambda}, or b) deploy multiple function instances of the same user in a single larger VM, i.e., the \emph{multi-container-per-VM} (N:1) model adopted by Microsoft Azure~\cite{azurefunctions}. 

The 1:1 model facilitates the immediate release of occupied resources when a function instance is reclaimed; it only requires
shutting down the VM that was hosting it. 
Unfortunately, this agility comes at the cost
of higher start-up delays, i.e., \emph{cold starts}~\cite{peeking,azuretrace}, 
and per-instance memory overhead,
as scaling up involves booting
a new (guest) Operating System (OS) 
and initializing the function's runtime. 
While a long line of research~\cite{catalyzer,vhive,faasnap,faascache,azuretrace} 
attempts to tackle these
performance pathologies
of the otherwise agile 1:1 model,  
the N:1 model inherently reduces the start-up cost
and the per-instance memory tax.  
It spawns new function instances in already running VMs, hence reusing the initialized guest OS and runtime
state of the VM. 
However, the N:1 model 
trades off resource allocation flexibility for cold-start performance. 
Naturally, it requires larger VMs, that will be able to accommodate the deployment of multiple concurrent function instances. 
However, these resources remain 
allocated --albeit idle-- even when the 
load is low, as function instances terminate and get reclaimed.

In this paper, we focus on the N:1 model and explore a solution for the agile adjustment of its VM resources.
Since compute resources can be (de)allocated on the fly~\cite{harvestserverless,hvm}, we focus on memory elasticity. 
As we show, 
sizing-up the VM memory is a relatively
cheap operation, but memory reclamation remains
a persisting problem. 
The state-of-practice interfaces for memory elasticity, i.e., memory ballooning~\cite{balloonvmware,resizingballoon} and vDIMM memory hotplugging~\cite{hotplugkernel}, reclaim memory unreliably and slowly,
consuming precious CPU resources, 
and potentially inducing fragmentation in the system
~\cite{virtio-mem,hotplugorballon,freepages}. 
virtio-mem~\cite{virtio-mem} is a state-of-the-art approach that improves upon both. 
However, we experimentally show that it still requires
multiple seconds to  reclaim 2 GiB of memory (\S~\ref{sec:eval:ubench:uspeed}). 
This is especially problematic for the FaaS use-case, as the bursty access patterns of FaaS 
require memory and CPU resource scaling in sub-second granularity~\cite{huawei-cold}. 

We identify the guest OS memory management as a key issue hindering VM memory reclamation performance.
\revision{The memory pages 
of processes running inside a VM (e.g., function instances) are typically spread 
across the guest physical address space and interleaved with pages from other processes, as modern OS\shepherd{es} allocate memory on demand and non-contiguously~\cite{capaging}.
Thus, when a process exits and releases its memory, 
the freed pages are commonly
scattered across blocks of guest memory 
that also hold occupied pages (Figure~\ref{fig:vanilla-zones}), for other processes or the OS.}
These pages need to be migrated before the blocks in question can be reclaimed by the host and 
these migrations dominate the performance of VM down-sizing~\cite{hotplugorballon}.
\emph{We argue that memory hot(un)plugging needs to be coupled with informed guest memory management to ensure fast and reliable VM memory reclamation.}

To this end, we design Squeezy: an extension
to the guest OS memory manager tailored for N:1 FaaS VMs.
Our insight is that the maximum memory per function instance is defined by the user. 
Squeezy is thus able to leverage the predictable memory requirements of FaaS to segregate the footprints of the N function instances that run inside the VM. 
It partitions the guest memory into 
fixed-sized chunks (partitions), based on the configured function memory, 
and optimizes their dynamic hot(un)plugging. 
It assigns a dedicated partition for each function instance in the system and extends the OS memory manager 
to serve each instance's allocation requests from its own dedicated partition. 
This way it eliminates the interleaving of the memory footprints of different instances, without breaking the fundamental property of on demand memory allocation.
We extend the hot(un)plug interface to make it aware of these partitions, i.e., to dynamically populate them during plug operations and to 
reclaim them instantly during unplug requests. 
With Squeezy, unplugging  
only involves removing the empty partitions of terminated function instances
without migrating any  
pages. 
By avoiding migrations, Squeezy also eliminates the memory bandwidth and CPU interference of unplugging, which can impact the performance of other concurrently running functions during VM down-sizing~\cite{mhvm}.

We integrate Squeezy with an OpenWhisk-based FaaS runtime~\cite{faasmem,openwhisk} 
that implements the N:1 model.
Squeezy interfaces with the runtime
to hot(un)plug partitions when it scales up and down the number of running instances based on the incoming load. 
Our evaluation reveals that Squeezy is an order-of-magnitude faster than  
state-of-the-art when reclaiming VM memory, achieving sub-second reclamation of multiple GiBs of memory while serving realistic FaaS load. 
\revision{When host memory becomes scarce, we show that Squeezy's agile VM resizing 
can keep tail latency bounded (10\% slowdown), while significantly reducing memory utilization.
Finally, we compare scaling up instances inside a Squeezy N:1 VM to using 1:1 microVMs and study the cold start performance and memory utilization.
}

\section{Background and Motivation}
\label{sec:motiv}

\subsection{Function Isolation Models}
\label{sec:motiv:isolation}
FaaS providers use two main models~\cite{rund} to securely deploy function instances into virtual machines: i) the multi-contai-ner-per-VM (N:1)~\cite{azure,azurefunctions}  and ii) the single-container-per-VM (1:1)~\cite{awslambda} model. 
The N:1 model scales up the number of function instances (containers) of the same user within the same VM to handle spikes in incoming load. The 1:1 model deploys instead each function instance in a lightweight VM, i.e., a microVM, and scales up the number of VMs. 
While both models are adopted by the industry ~\cite{azure,awslambda}, each presents its own trade-offs with respect to i) cold start delays, ii) resource sharing, 
and iii) resource allocation. 

\noindent\textbf{Cold starts.} 
During scale up events, the 1:1 model 
incurs the overhead of booting a new microVM, 
hence penalizing cold start execution. 
For this reason, providers typically opt 
for keeping idle VMs alive~\cite{azuretrace,faascache}, in order to avoid cold start costs, trading
tied down idling resources for 
sustainable performance~\cite{huawei-trace,huawei-cold,azuretrace,peeking}. 
While a long line of research attempts to optimize the VM cold-start overhead~\cite{vhive, faasnap, peeking, fireworks,proghorn}, starting a new VM is in principle a costlier operation than creating a new container~\cite{survey,agile} -- which is how the N:1 model scales up instances. 
The N:1 model also incurs less overhead for the runtime initialization, during cold starts, as the runtime dependencies of new function instances are typically already cached in the shared VM, e.g., the container root file system could already be cached by the guest OS page cache. 
\revision{In our experiments we find that cold start execution is 1.6x faster in the N:1 model (\S\ref{sec:eval:comparison}).}

\begin{mybox}
{\it 
\shepherd{The multi-container-per-VM model (N:1) reduces the microVM boot overhead of cold starts (Figure~\ref{fig:1-1}).}
}
\end{mybox}

\noindent\textbf{Resource sharing.} 
The 1:1 model also has higher memory costs, as it requires a separate guest OS and function runtime per function instance. 
As already mentioned earlier, the N:1 model instead shares the OS and runtime state among the instances that are co-located within the same VM. 
\revision{In our experiments we find that the 1:1 model increases a new instance's footprint by 2.53x on average (\S\ref{sec:eval:comparison}).} 

\begin{mybox}
{\it
The multi-container-per-VM model (N:1) has a lower memory tax per function \revision{(Figure~\ref{fig:1-1}).}
}
\end{mybox}

\begin{figure}[h]
\centering
  \vspace*{-5.0mm}
  \begin{subfigure}[b]{0.9\linewidth}
        \centering
   \includegraphics[width=0.99\linewidth]{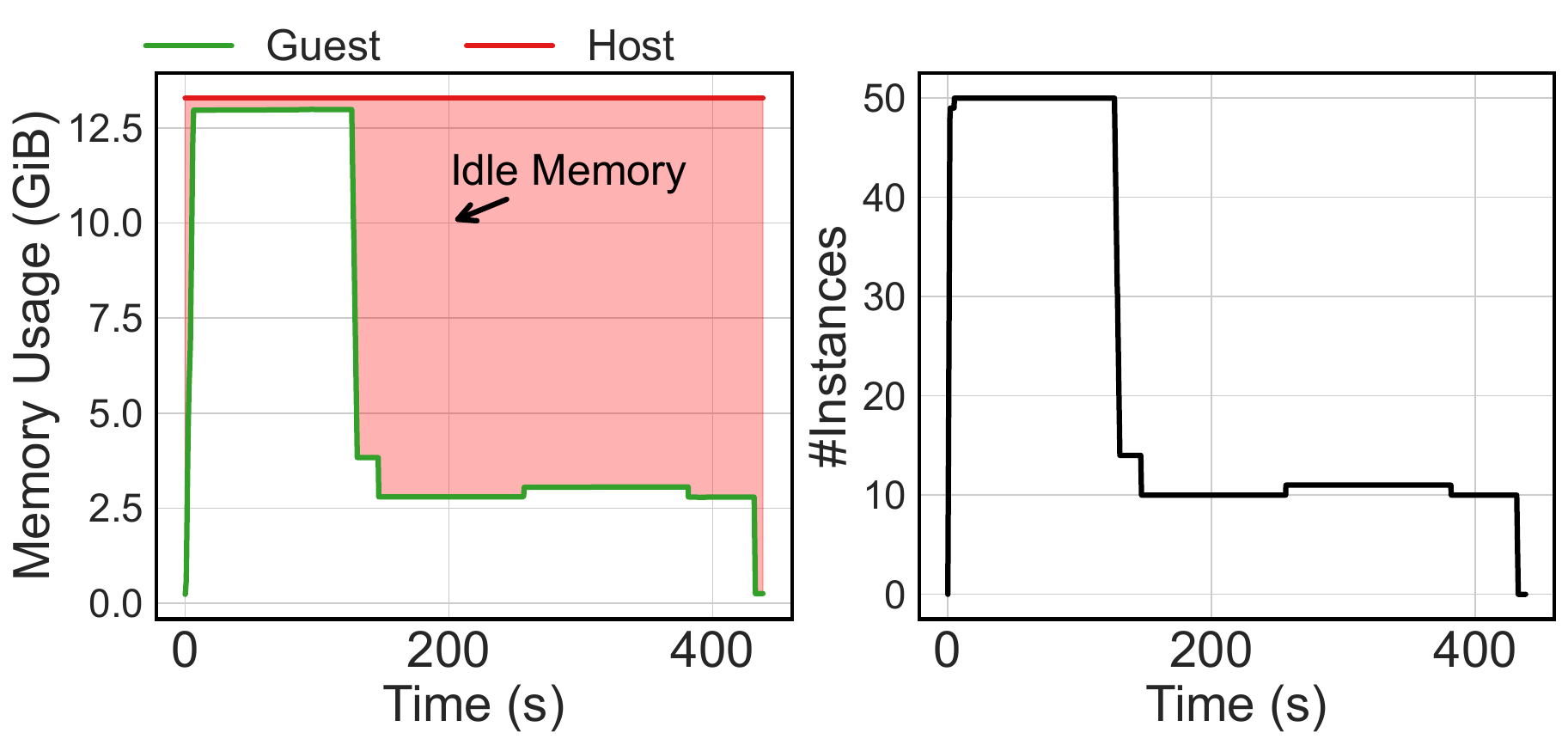} 
   \end{subfigure}
   \vspace*{-2.0mm}
   \caption{\revision{The N:1 model reserves memory for N instances even when the load is low and a large number of concurrent instances are reclaimed by the FaaS runtime.}}
   \label{fig:motiv:mem}
  \vspace*{-2.0mm}
\end{figure} 

\noindent\textbf{Resource allocation.} 
In the 1:1 model, the resources allocated per instance (microVM) follow the target function's resource limits as defined by the user~\cite{huawei-trace}.
Thus, when the provider decides to scale down function resources, it can shut down a microVM and instantly release a predictable amount of memory and CPU shares in the host system. 
On the other hand, in the N:1 model, the resource reservation
is subject to the concurrency factor, i.e., the number of function instances provisioned to be deployed concurrently within the same VM. For this model, VMs are typically provisioned with enough memory to accommodate multiple (N) instances.
However, as the number of required instances varies over time, e.g., based on the incoming load, this model frequently results in memory waste~\cite{rund,survey,pond,mhvm}. 

\revision{
Figure~\ref{fig:motiv:mem} shows the memory utilization, i.e., the allocated memory, measured inside a VM (guest), configured for 50 concurrent function instances (50:1), and the host system, when we serve a bursty trace of requests, based on real-world load~\cite{azuretrace,harvestserverless}, using a N:1 FaaS runtime (\S\ref{sec:meth}).
The figure also reports the number of running instances over time.
We observe that the memory utilization in the guest increases as the FaaS runtime scales up the number of concurrent instances to serve the spikes in the incoming load.
Later, when the load drops, the runtime evicts the majority of these instances and the memory utilization decreases. However this elasticity is not reflected in the host system, where idle memory remains allocated, matching the maximum concurrency factor that the VM reached since it booted.
}

\begin{mybox}
{\it
The multi-container-per-VM model (N:1) ties down idle resources when the load is low \revision{(Figure~\ref{fig:motiv:mem}).}
}
\end{mybox}

\begin{figure}[h]
\centering
  \begin{subfigure}[b]{0.9\linewidth}
        \centering
        \includegraphics[width=0.99\linewidth]{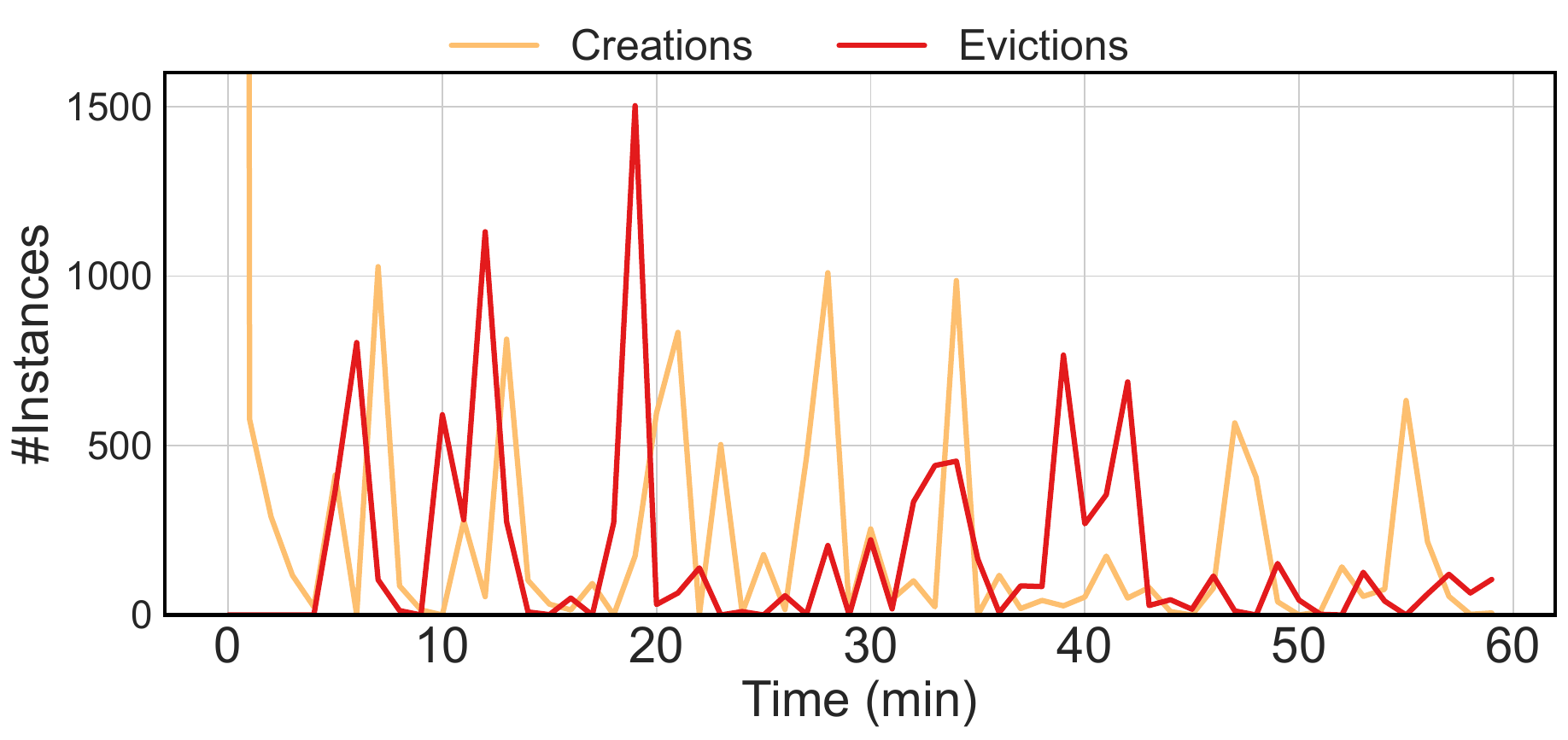} 
    \end{subfigure}
    \vspace*{-2.0mm}
   \caption{\revision{An analysis of the invocation pattern of the 10 most popular functions in the Azure production traces~\cite{azuretrace} shows that thousands of instances are created and evicted per minute. Mechanisms that enable the dynamic and agile resizing of N:1 VMs are essential to save resources.}}
   \label{fig:motiv:instances}
   \vspace*{-5.0mm}
\end{figure}

We consider this rigid resource over-provisioning~\cite{pond,mhvm,fcgh} one of the major drawbacks of the N:1 model that may overshadow its benefits. 
\revision{To put this into scope, Figure~\ref{fig:motiv:instances} depicts 
the number of instances that are created and evicted per minute for the 10 most popular functions in the Azure production traces~\cite{azurefunctions}, when we simulate a random hour in the trace and assume that idle instances are evicted after 5 inactive minutes. We observe that thousands of instances can be scaled up and down per minute.
Reclaiming memory from the evicted ones and re-distributing it to the newly created instances is essential for resource efficiency.}
A natural way to address the problem is to reclaim memory and CPU resources at runtime. While CPU reclamation is well studied ~\cite{rund,harvestserverless,azurespot,awsspot,hvm},
the efficient dynamic resizing of VM memory remains an open challenge, especially under the tight latency requirements of FaaS~\cite{rund}.

\subsection{Dynamically Adjusting VM Memory}
\label{sec:motiv:resize}

The state-of-practice and art techniques to resize a VM's memory are a) ballooning~\cite{xen,vmware-balloon} and 
b) virtio-mem~\cite{virtio-mem} vDIMM memory hotplugging~\cite{virtio-mem,hotplugkernel}.
Memory ballooning implements a paravirtualized interface, 
where a guest driver (controlled by the hypervisor) allocates (inflation) and de-allocates (deflation) guest memory pages to scale down and up its memory resources. 
Our results (\S\ref{sec:eval:ubench}) corroborate past studies~\cite{virtio-mem} that ballooning suffers from 
high scaling overhead as it reclaims and releases memory to the host at the granularity of a page.

Virtio-mem~\cite{virtio-mem} dynamically adjusts VM memory by emulating the hot(un)plugging of physical DIMMs in the hypervisor. 
It exposes a new device, i.e., the paravirtualized DIMM,
sliced in small slots each capable of independent (un)plugging for agile VM memory adjustment. 
While faster than ballooning, our study corroborates
past findings~\cite{mhvm,hotplugorballon,hubballon}
that virtio-mem still suffers from high OS overhead.

\label{sec:background:linux}
Virtio-mem relies on the native Linux mechanisms to hot(un)plug memory. 
Linux manages memory in the granularity of pages (i.e., 4KiB), but adds and removes ((un)plugs) memory in the granularity of blocks, i.e., 128MiB for x86.

\noindent{\textbf{Hotplugging a memory block}} 
involves two major steps: 
a) add the memory block to the OS (\emph{hot-add}) and b) expose it to the OS allocator (\emph{online}). \emph{Hot-add} updates the OS metadata for physical memory, such as the Linux memory map array (\emph{memmap}), 
and \emph{online} releases the new block's pages to the OS allocator, exposing the newly-plugged memory as usable.
\revision{
Our results (\S\ref{sec:eval:faasazure}) indicate that 
hotplug is a relatively cheap operation, 
especially for the number of blocks
that FaaS functions typically require (\S\ref{sec:meth}).
}

\noindent{\textbf{Hotunplugging a memory block}}
involves two major steps as well: 
a) retract the pages of the block from the OS allocator  (\emph{offline}) and b) remove the block from the OS by destroying all the corresponding metadata (\emph{hot-remove}). 
The first step dominates unplug performance as it 
commonly involves migrating occupied pages.

\emph{Page migrations} are indirectly 
induced by the guest OS memory manager.
Memory that is hot-plugged is typically added to a special zone of the OS memory allocator, i.e., the \texttt{ZONE\_MOVABLE} memory zone.
This special zone 
separates movable 
and non-movable pages in the system
in an attempt to guarantee that blocks in this zone can be offlined~\cite{contiguitas,making}. 
The memory manager uses it to serve all allocation requests for data that in principle can be migrated, e.g., user-space allocations and file system caches (i.e., page cache). The memory blocks that populate the zone can be offlined  
as soon as all their occupied pages are migrated to other free pages / blocks in the system.
As the OS commonly interleaves the footprint of different processes~\cite{ranger,capaging}, such expensive migrations 
are frequent when attempting to unplug the memory of a process after it terminates. 

\begin{figure}[h]
\centering
  \begin{subfigure}[b]{0.46\linewidth}
        \centering
   \includegraphics[width=0.99\linewidth]{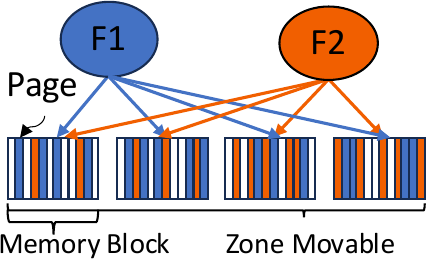} 
   \caption{Vanilla Linux}
   \label{fig:vanilla-zones} 
   \end{subfigure} 
   \hfill
    \begin{subfigure}[b]{0.46\linewidth}
        \centering
   \includegraphics[width=0.99\linewidth]{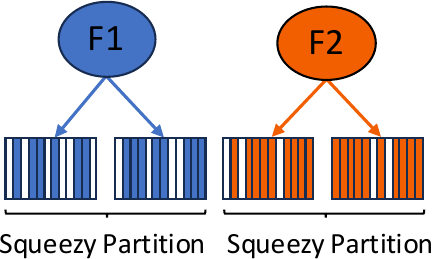} 
   \caption{Squeezy}
  \label{fig:squeezy-overview} 
   \end{subfigure}
\caption{ Linux interleaves processes footprint within physical memory blocks. When F2 exits, the F1 pages that are co-located in the same blocks have to be migrated to reclaim memory. Squeezy partitions processes footprints.}
\label{fig:vanilla-zones} 
\end{figure}

Figure~\ref{fig:vanilla-zones} depicts a VM with 2 running function instances. The OS has mapped their virtual addresses with pages only from \texttt{ZONE\_MOVABLE}, but their footprints are interleaved across multiple memory blocks of the zone. Linux allocates memory lazily, at the page granularity (4KiB or 2MiB), when the processes access their pages for the first time (page fault). To serve a  fault, the OS allocates any available page, hindering contiguous mappings, scattering the workload's footprint
and interleaving it with other processes.

This interleaving eventually penalizes memory unplugging. 
For example, in Figure~\ref{fig:vanilla-zones}, when the instance \emph{F2} terminates and releases its memory pages,
the host may request to reclaim the released memory. 
The guest will try to isolate (offline) memory (e.g., via virtio-mem) equal to the size of the terminated instance by iterating over the system's memory blocks.
Due to the interleaving with the F1's footprint, 
most blocks will be at least partially occupied, thus the guest will have to a) migrate some pages to isolate enough memory blocks, and b) then offline and unplug them. 
\revision{
In Section~\ref{sec:eval:ubench:uspeed} we measure that these migrations slow down 
unplugging by 61.5\% (on average across different unplug memory sizes). These migrations can also penalize the performance of other running processes in the VM as they consume precious CPU cycles~\cite{virtio-mem} (\S\ref{sec:eval:faasazure} -- Figure~\ref{fig:cpuusage}).
}

\begin{mybox}
{\it
Page migrations hinder VM memory reclamation performance and can penalize other processes running inside the down-sizing VM \revision{(Figures ~\ref{fig:ubench:latency} and ~\ref{fig:interference}).}
}
\end{mybox}

\revision{ 
Finally, another source of overhead in memory unplugging  stems from the 
\shepherd{unnecessary} zeroing of memory blocks. 
When unplugging memory, virtio-mem uses generic routines of the OS memory allocator in order to reserve (\emph{isolate}) the memory to be \emph{offlined} from the rest of the system. 
The Linux kernel is commonly configured to zero pages upon allocation~\cite{initonalloc}, as a security hardening measure.
This configuration leads to the 
\shepherd{unnecessary} zeroing of the about-to-be-unplugged memory, as the allocator routines are oblivious to the hot(un)plugging operations in progress, and they default to zeroing-out the offlining pages. 
We find that \emph{zeroing-out memory blocks} that are about to be offlined accounts for 24\% of unplug operations latency on average (Figure ~\ref{fig:ubench:latency}). 
}

\section{Squeezy Overview}\label{squeezy-overview}
\label{sec:overview}
\label{sec:design}
We observe that the key reason that hinders VM memory reclamation performance 
is that the existing hot(un)plug interfaces attempt to blindly reclaim guest OS memory. 
For example, virtio-mem~\cite{virtio-mem} needs to scan the guest physical memory linearly and migrate occupied pages to isolate and remove memory blocks. 
While it is hard to identify and isolate free memory on general purpose VMs, e.g., VMs hosting multiple long-running workloads with working sets that vary over time and input, we observe that memory usage is far more predictable for the serverless use-case. The occupied and free memory inside a VM can be identified and managed at the granularity of the deployed function's memory resource limits, as defined by the user. 
\emph{However, to be able to hot-add and hot-remove such coarse-grained blocks fast, avoiding migrations,
special mechanisms are necessary to guarantee the isolation of instances footprints.}

The goal of this work is to provide such mechanisms to dynamically and efficiently resize N:1 FaaS VMs while reducing the memory waste of the original N:1 model (\S\ref{sec:motiv:isolation}).
We build on existing state-of-the-art memory hot(un)plugging interfaces, i.e., virtio-mem, and attempt to address their performance pathologies, as identified in Section~\ref{sec:motiv:resize}, under the scope of the FaaS use-case. 

To that end, we design Squeezy, an extension to the guest OS memory manager, to efficiently manage the memory of serverless functions, enabling its rapid reclamation.
Aligned to our analysis in Section~\ref{sec:motiv}, we set the following main goals:

\noindent{\textbf{Resource elasticity.}}
Squeezy should be able to 
track the memory requirements of the total number of alive function instances running inside the same VM, scaling up and down its memory resources accordingly. 
Squeezy should also be fast enough to cope with the bursty load of serverless functions that frequently requires the up- and down-scaling of resources in (sub)second intervals~\cite{huawei-cold}.

\noindent{\textbf{Resource sharing.}}
Squeezy should allow the sharing of runtime state, e.g., libraries, across the function instances that run in the same VM to preserve the memory saving benefits of the N:1 model. 

\noindent{\textbf{Minimal interference.}}
Squeezy should not affect the performance of other running function instances in the VM. 

To achieve them, Squeezy introduces two mechanisms:
\begin{itemize} [leftmargin=2ex, noitemsep, topsep=1pt]
\item  It partitions guest memory between running function instances to isolate their footprints.
\item It makes the OS hot(un)plug interface (i.e., virtio-mem) aware of these partitions, to enable their fast 
and efficient reclamation when instances terminate.
\end{itemize}

\noindent\textbf{Squeezy partitions.} 
Squeezy leverages the pre-defined memory requirements of function instances to partition the available guest memory of a N:1 VM into N fixed-sized chunks (partitions). 
Each chunk can host a single function instance. 

Squeezy memory cannot be used for generic allocations; it is reserved and excluded from the main OS allocation path. Instead, an instance explicitly asks to be backed by Squeezy memory, using a dedicated API described in Section~\ref{sec:implementation}. In that case, it is mapped to a Squeezy partition and all its allocations (i.e., anonymous memory) are served from it. Once a partition is assigned to an instance, it is locked until the instance terminates. Figure~\ref{fig:squeezy-overview} shows how different instances are isolated into different memory partitions under Squeezy.

Squeezy partitions are initially empty. 
When the serverless runtime decides to scale-up the number of function instances running inside the N:1 VM, e.g., due to a load spike, it triggers a plug event. This event will populate a Squeezy partition with the hotplugged memory and the runtime will 
spawn and assign a new function instance to this partition (\S\ref{sec:runtime}). 

\revision{
\noindent\textbf{Physical memory allocation.} 
Similarly to normal VM memory hotplugging with virtio-mem, Squeezy does not allocate any physical memory on plugging. The hotplugged guest physical memory corresponds to a virtual memory range of the virtual machine monitor (VMM). Physical memory pages will be allocated on demand, when the Squeezy-enabled function in the guest actually touches the hotplugged memory for the first time, via nested page faults. Thus its static partitioning does not increase memory usage; each instance will occupy only the amount of memory it touches and not the entire Squeezy partition size.
}

\noindent\textbf{Distinguishing shared and private allocations.}
To preserve the memory savings of the N:1 model, Squeezy employs a shared partition per VM to back the file mappings of all instances. File mappings typically correspond to libraries and runtime dependencies shared across instances, i.e., they are instantiated once in memory and mapped multiple times. This partition is pre-populated at boot time for each VM based on the characteristics of each function. The anonymous mappings of the instances are backed by private partitions as discussed above.

\noindent\textbf{Partition-aware unplugging.}
As instances run isolated into Squeezy partitions, unplugging the memory of a terminated instance involves zero migrations, thus minimal interference, under Squeezy. Unplug operations are triggered by a serverless runtime (\S\ref{sec:runtime}), when instances inside a VM are reclaimed due to a drop in the incoming load. Squeezy extends the unplug interface of the OS to directly identify the Squeezy partitions that are emptied when the instances terminate, and instantly offlines them.

\section{Squeezy Implementation}
\label{sec:implementation}

\noindent We implement Squeezy in Linux v6.6, extending: i) the (guest) OS memory manager and ii) the virtio-mem hot(un)plug guest driver (\S\ref{sec:design:squeezymech}). We also integrate Squeezy to a FaaS runtime to dynamically resize N:1 VMs (\S\ref{sec:runtime}).

\subsection{Squeezy OS Mechanisms}
\label{sec:design:squeezymech}

\noindent\textbf{Extending the OS memory manager.}
As described in Section~\ref{sec:overview}, Squeezy partitions the guest physical memory into chunks, that will host non-interleaved function instances. The size of each Squeezy partition must match the memory requirements of the function to be deployed in the specific VM under the N:1 serverless model. This parameter is known when the VM is set-up by a serverless runtime, as the memory resource limits of each function are defined by the user (\S\ref{sec:runtime}). Thus we configure the partition size as a boot parameter for the guest OS. 

We implement Squeezy partitions as different zones (\emph{zone structs}), similar to \texttt{ZONE\_MOVABLE}, in the Linux physical memory manager. 
Each zone  
represents the guest physical memory region of a Squeezy partition and is stored in the per-NUMA node zonelists of the kernel (\emph{node\_zones}).
Each partition is uniquely identified by its \emph{partition id}. 

At boot time, we create $N$ such Squeezy zone structures, that initially link to empty partitions.
This sets the maximum concurrency that can be supported by the VM, i.e., only $N$ function instances can be concurrently deployed at any point in time. 
We use the term \emph{concurrency factor} for $N$ and expose it as a boot parameter to be set by the serverless runtime.
\shepherd{We note that the maximum memory requirements per instance, i.e., partition rated size, are identical in this design, as each instance refers to the same function or to functions with similar memory requirements.}
\revision{We further discuss the implications of 
choosing $N$  
in Section~\ref{subsec:discussion}.
Note that unlike an over-provisioned VM, this design pre-sets the maximum concurrency but does not pre-allocate the corresponding memory resources. 
\shepherd{The $N$ Squeezy partitions are initially empty (at boot time); they are not backed by physical memory pages, as discussed in Section~\ref{sec:overview}.}
They are instead populated and emptied dynamically by plug and unplug operations, triggered by the FaaS runtime (\S\ref{sec:runtime}).
}

Squeezy populates only one partition at boot time, the \emph{shared Squeezy partition}, that is dedicated to store shared runtime and language dependencies across concurrently running function instances in the VM. 
Its size is also a boot time parameter set by the serverless runtime
We elaborate in Section~\ref{sec:runtime} how the FaaS runtime decides what size to use.

\noindent\textbf{Plugging a Squeezy partition.}
The Squeezy partition plugging starts when the hot(un)plug driver inside the guest, i.e, virtio-mem, receives a plug request from the hypervisor. 
The driver becomes aware that a specific range, corresponding to the partition(s) size, of its managed memory has been plugged by the hypervisor and uses the kernel memory onlining interfaces accordingly. 
We intercept the onlining process in order to instruct the kernel to correlate the memory range with the corresponding Squeezy partition(s), populate the free pages of the Squeezy partition and notify the users of the interface, as explained in the next paragraph.

\noindent\textbf{The Squeezy interface.}
We design a system-call based interface for Squeezy, which allows the assignment of populated Squeezy partitions to function instances.
The interface allows the calling process to request the OS to serve its memory allocations via Squeezy partitions.
For each such request, we scan the list of Squeezy partitions (zonelist) in order to find an available (empty) Squeezy partition. 
Upon success, the partition is marked as reserved. 
Per-partition locks are used to avoid race conditions for concurrent requests. 

Linux uses a memory descriptor (\emph{mm\_struct)} to represent each process's address space.  
We add a new field in the memory descriptor to store the partition id that is assigned to each Squeezy process. 
This is then used on the memory allocation path,
in order to only allocate pages from the specific partition for the process. 

\revision{
\noindent\textbf{Squeezy waitqueue.} 
Squeezy decouples the onlining of Squeezy partitions via (hot)plug events, from their assignment
to processes (function instances) via the Squeezy syscall interface. 
While the FaaS runtime orchestrates and effectively couples these events (\S\ref{sec:runtime}), i.e., it issues plug operations which populate Squeezy partitions in tandem with corresponding Squeezy-enabled function instance creations, 
these two events happen asynchronously. 
\shepherd{There is therefore a chance that requests for Squeezy partitions may occur before the Squeezy zone has been fully populated and brought online.}
We use  
a \emph{waitqueue} for the 
synchronization of partition assignment requests. 
When a process requests to be assigned to a Squeezy partition  
we first check if a Squeezy partition is populated and free. 
If not  
the requesting process is placed in the afore-mentioned \emph{waitqueue}, until a plug operation that populates a Squeezy partition completes.
We note that the setup of the OS sandboxing mechanisms (e.g., cgroups, network) can proceed in parallel with the plugging event, i.e., before the process (function instance) is assigned to a Squeezy partition.
}

\noindent{\textbf{Squeezy memory allocation.}} As a Squeezy process starts executing inside the VM, its virtual memory accesses trigger the lazy allocation of guest physical memory through page faults. Squeezy intercepts the faults and implements a dedicated handler that allocates pages from the Squeezy partition that is assigned to the process, instead of allocating pages from generic and shared OS zones (e.g., from \texttt{ZONE\_MOVABLE}). 
This guarantees the confinement of the footprint of each Squeezy process within a single partition. 

Since the function instances running in a N:1 FaaS VM correspond to the same function (\S\ref{sec:motiv}), they share the same container root file system and runtime dependencies. 
To achieve the design goal of preserving the sharing of language and runtime dependencies among the function instances running in the VM (\S\ref{sec:overview}), Squeezy distinguishes between faults for anonymous and file mappings. 
For the first, it allocates pages from the process-assigned Squeezy partition, as described above. 
For the latter, it uses the system-wide \emph{shared Squeezy partition}, as described in previous paragraphs. 
It faults-in each file page once and subsequently maps it to the address space of any Squeezy process that touches it upon a fault. 
Essentially, this design implements a file caching layer for all function instances via a dedicated shared Squeezy partition. 

As discussed above, Squeezy applies the user-set memory limits of each function via the Squeezy partition size.
If a function tries to allocate more memory than the size of the Squeezy partition,
OS mechanisms (e.g, the OOM Killer) are triggered to kill the Squeezy process and prevent violations of partition isolation.

\noindent\textbf{Handling \texttt{fork()}.}  
When a Squeezy process calls the \texttt{fork()} system call to create a new process, we assign the child to the parent's Squeezy partition. 
Thus we co-locate all threads and processes of a function instance on the same Squeezy partition.  
To handle this case, we add a partition reference counter (\emph{partition\_users}) for each Squeezy partition, to track the number of concurrent processes (\emph{mm\_structs}), that each partition is assigned to.  
When the \emph{partition\_users} reaches zero, the partition is no longer in use, thus we mark it as free, hence reclaimable by virtio-mem.

\noindent\textbf{Unplugging a Squeezy partition.}
The unplug operation starts upon receiving an unplug request from the hypervisor. 
The hot(un)plug driver, i.e, virtio-mem, inside the guest, keeps track of free Squeezy partition(s) via their reference counter (\emph{partition\_users}), as described above. 
When the driver receives an unplug operation from the hypervisor, it immediately offlines and removes a free partition without migrating any of the pages, as the partition is guaranteed to be empty. 
The serverless runtime coordinates the hypervisor unplug operations with the function recycle operations.
It can thus know the exact number of free Squeezy partitions in the VM.
The virtio-mem driver then informs the hypervisor that the memory blocks corresponding to the unplugged Squeezy partition(s) have been removed. 
The hypervisor immediately releases the pages to the host using the \emph{madvise()} system call to mark them as not needed (\texttt{MADV\_DONTNEED}). 

\revision{
As discussed in Section~\ref{sec:motiv}, virtio-mem might trigger the \shepherd{unnecessary} zeroing of pages that are to be unplugged, due to the obliviousness of the Linux memory allocator to the unplug operations in progress. 
We thus modify the Linux memory allocator to be hot(un)plug aware and skip the zeroing of the pages that are about to be unplugged. 
This memory will eventually get zeroed out later, when it is re-allocated, either by the host or by another VM. 
}

\begin{figure}[h] 
 \centering
\includegraphics[width=0.99\linewidth]{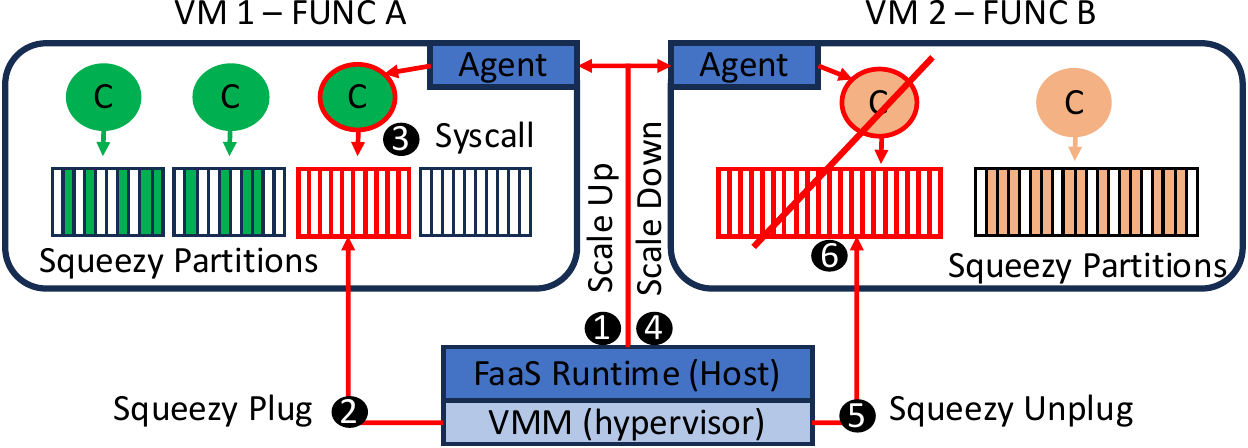} 
\caption{Squeezy integration into a serverless runtime}
\label{fig:runtime} 
\vspace{-3ex}
\end{figure}

\subsection{Squeezy Integration Into a FaaS Runtime}
\label{sec:runtime}

Squeezy is an OS mechanism designed for the rapid reclamation of VM memory resources of serverless function instances when they terminate. 
A serverless runtime can leverage the Squeezy interface in order to (un)plug and (de)allocate Squeezy memory as it scales up and down function instances in the N:1 serverless model. Figure~\ref{fig:runtime} shows how Squeezy can be integrated into such a runtime. 
For simplicity, we show the scaling up / down workflow for only one function instance.
However, the same applies when the runtime concurrently scales up or down multiple function instances to respond to spikes in the incoming load. In Section~\ref{sec:eval}, we evaluate a Squeezy-aware FaaS runtime based on OpenWhisk.

\noindent\textbf{Scaling up.}
Figure~\ref{fig:runtime} (left) shows the steps that take place when the runtime decides to spawn a new function instance to scale up resources for function A. The runtime sends a request to to the agent that runs inside the VM to scale up function instances~\circleint{1}. 
It also employs the hypervisor in the host to send a plug request to the guest, asking to add memory equal to the requirements of the instance, i.e., the memory size pre-defined by the user. 
The Squeezy plug path is triggered, and the guest populates a Squeezy partition 
~\circleint{2}. 
The agent then spawns a new instance using the \emph{Squeezy syscall interface}~\circleint{3}.
A new container is created and starts executing with all its private allocations being served by and isolated in the just-plugged Squeezy partition. 

\noindent\textbf{Scaling down.}
Figure~\ref{fig:runtime} (right) shows the steps that take place when the runtime decides to evict a function instance to scale down resources for function B. 
The agent that runs inside the VM shuts down the function container and informs the runtime for the successful down-scaling~\circleint{4}. 
The runtime then employs the hypervisor in the host to send an unplug request to the guest asking to reclaim memory equal to the memory freed by the recycle~\circleint{5}. 
The Squeezy unplug path is triggered in the guest, which i) identifies the empty Squeezy zone of the recycled container, ii) directly removes it without any migrations, and iii) notifies the hypervisor~\circleint{6}.

\noindent\textbf{VM creation.} 
As discussed earlier, when the runtime creates a VM that will host Squeezy function instances it must declare during boot i) the Squeezy private partition size, matching the memory resource limit set for the function, ii) the shared Squeezy partition size, matching the size of the runtime and language dependencies of the function, and iii) the maximum number of function instances that will be deployed in the VM (concurrency factor $N$). 

\vspace*{-2.0mm}
\section{Methodology}
\label{sec:meth}

\subsection{Experimental Setup}
\noindent\textbf{Hardware setup.}
We use a dual-socket 40-core Intel
Xeon E5-2630 server
with 128GiB of memory per socket
as a host. For the dynamically resized
N:1 VMs of the FaaS experiments, 
we set the number of vCPUs per VM based 
on the CPU shares of the target function (Table~\ref{tab:functions}) and the max concurrency factor (N) of every experiment. 
To minimize jitter, we use a single NUMA node, pin each vCPU on a single core, and set the core frequency to 2.2GHz. 
We also enable the Transparent Huge Page (THP) mechanism in the host.

\noindent\textbf{System software.}
We implement Squeezy on Linux Kernel 6.6.30
in $\sim$~800 Lines of Code (LOC). 
We use the Cloud Hypervisor (v38.0) as the virtual machine monitor (VMM) for the N:1 VMs as it includes a well tested Rust implementation of the virtio-mem~\cite{virtio-mem} interface. 
We use microVMs for the evaluation of the 1:1 model.

\begin{table}[h]
\vspace*{-2.0mm}
\centering
\footnotesize
\begin{tabular}{cccc}
 \textbf{Function} & \textbf{Description} & \textbf{vCPU shares} & \textbf{Memory (MiB)} \\
 \midrule
 Cnn & JPEG classification& 1.0 & 768\\
 Bert & ML inference & 1.0 & 1536\\
 BFS & Breadth-first search & 1.0 & 768 \\
 HTML & Web service & 0.25 & 768 \\
 \end{tabular}
 \caption{Serverless functions used in the evaluation and their assigned resource limits per instance.
 }
 \label{tab:functions}
 \vspace*{-8.0mm}
\end{table}

\noindent\textbf{Workloads.}
We use the memhog~\cite{memhog} microbenchmark and the 
FaaS 
functions of 
Table~\ref{tab:functions}
for our evaluation. 
They consist of a function from FunctionBench~\cite{functionbench} (CNN) and 
three real-world functions from~\cite{faasmem}
(HTML, BFS, and Bert).
\shepherd{These functions are sufficiently memory-intensive to assess Squeezy’s allocation and reclamation mechanisms realistically. They also represent both main memory allocation types: anonymous memory (e.g., BFS) and file-backed page cache (e.g., HTML, Bert, CNN). We aim to show that Squeezy performs well under high memory demand, and we expect similar benefits in scenarios with a higher number of small-sized functions.}
The table also summarizes the vCPU and memory limits that we use for the function instances (containers)~\cite{faasmem}.
The memory limits are tailored
for concurrency N=1 (Figure ~\ref{fig:1-1:memory}).

\subsection{Evaluation Scenarios}
\label{sec:meth:eval}

\noindent\textbf{Reclaiming VM memory.}
We benchmark the downsizing of a N:1 VM running memhog. We study a) memory reclamation latency and b) CPU utilization. We compare Squeezy to a) vanilla virtio-mem memory unplugging (\emph{virtio-mem}) \revision{and b) to reclaiming memory using the balloon driver (\emph{balloon}).}

\noindent\textbf{Integration into a FaaS runtime.} 
We study the performance of a FaaS OpenWhisk-based autoscaler~\cite{faasmem} that uses dynamically resized N:1 VMs. 
We evaluate a) the achieved memory reclamation throughput, b) cold start delays and c) the CPU interference 
when the runtime serves requests from a production trace~\cite{harvestserverless} 
and there is abundance of memory in the host system.
\revision{
We also study performance
when we 
restrict the available memory in the host to emulate the spawn and reclaim patterns of a
large scale end-to-end experiment (Figure~\ref{fig:motiv:instances}).
For this experiment, 
the memory reclamation speed of
scale down events impacts 
end-to-end performance,
as VMs have to wait for 
memory to be freed to scale 
up their number of instances.
We compare Squeezy performance
to a) vanilla virtio-mem and 
b) a version of virtio-mem that incorporates optimization techniques from HarvestVM~\cite{harvest} (proactive reclaim, buffering).
}

\noindent\revision{
\textbf{1:1 vs. N:1 model performance trade-offs.} 
Finally, we compare the cold start delays 
and the memory usage of the N:1 model, 
implemented with dynamically resized VMs 
via Squeezy, to using 1:1 microVMs.  
}

\section{Evaluation}
\label{sec:eval}

\subsection{Reclaiming VM Memory}
\label{sec:eval:ubench}
We first evaluate memory reclamation
from a N:1 VM that hosts 
multiple instances of memhog~\cite{memhog}. 
Memhog repeatedly (de)allocates 
chunks of memory of fixed size
and thus stresses 
the CPU and memory usage of the VM. 
We compare Squeezy memory hot-unplugging to 
vanilla virtio-mem unplugging \revision{and to balloon
inflation.} For Squeezy, we deploy each
memhog instance
within a plugged squeezy partition.

\begin{figure}[t!] 
 \centering
 \begin{subfigure}[b]{0.999\linewidth}
 \includegraphics[width=0.99\linewidth]{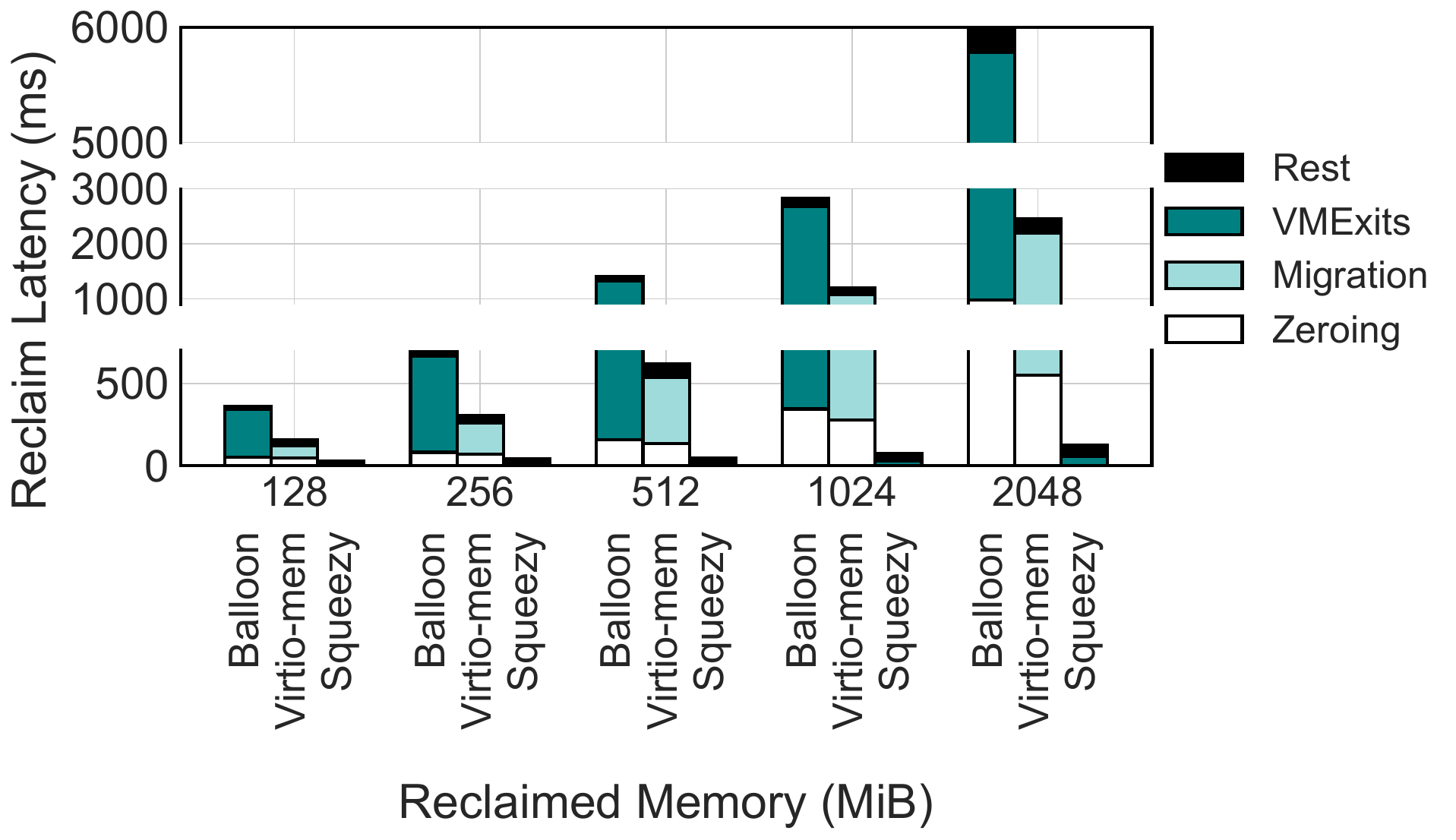} 
 \end{subfigure}
\vspace*{-2.0mm}
\caption{Average latency (ms) to reclaim memory of different sizes from a guest with memhog-loaded CPUs.}
\label{fig:ubench:latency} 
\vspace*{-5.0mm}
\end{figure}

\subsubsection{Reclamation Latency}
\label{sec:eval:ubench:uspeed}
For this set of experiments we spawn 32
memhog instances on a 32:1 VM.
We tailor the total VM memory size 
so that the 32 instances fully 
occupy its entire memory. 
We let the instances execute 
for a warm-up period and then
we kill them iteratively.
The host reclaims VM memory that equals 
the killed instance's memory size on every step. 
We report the average latency of the 32 
reclamation steps. 
We repeat the same experiment while
increasing the memhog memory size. 
All considered sizes are representative of the
memory limits of FaaS functions. 

Figure \ref{fig:ubench:latency} shows the reclamation latency achieved by the different methods
per memory size. 
\revision{
We break down the latency 
to the costs of 
a) page zeroing (guest), 
b) page migration (guest), 
c) serving VM exits (host)
and d) the rest.  
}

\revision{
As expected, ballooning is the less performant interface and is dominated by the costs of serving VM exits (81\% on average). To reclaim memory, the guest balloon driver allocates (reserves) guest physical pages and reports them back to the hypervisor (VM exit) which releases them
to the host (balloon inflation). 
The interface operates at the granularity of a page, 
thus when the reclaimed memory size 
increases the overhead of the technique explodes. 
}

Virtio-mem hot-unplugging is 
2.34x faster than ballooning (on average), as it reclaims memory at larger chunks, i.e., in 128 MiB memory blocks, 
and thus eliminates VM Exit overheads. 
However, it still requires 617 ms to reclaim 512 MiB of memory and almost 2.5 seconds to reclaim 2 GiB. 
\revision{
The overhead stems from the migration of occupied pages 
per reclaimed memory block (61.5\% on average) and their zeroing (24\% on average). 
}

Squeezy eliminates both the page migration and the zeroing overhead 
of virtio-mem, and is 10.9x faster in reclaiming memory (on average). It only requires 127 ms to reclaim 2 GiB of memory. Its partitions isolate a process's footprint  
in contiguous memory blocks; when the process exits,  
the blocks can be reclaimed instantly without any migrations
as there are no occupied pages by other processes.
Squeezy also defers zeroing the blocks to the host.

\begin{figure}[t!] 
 \centering
    \begin{subfigure}[b]{0.85\linewidth}
    \centering
   \includegraphics[width=0.99\linewidth]{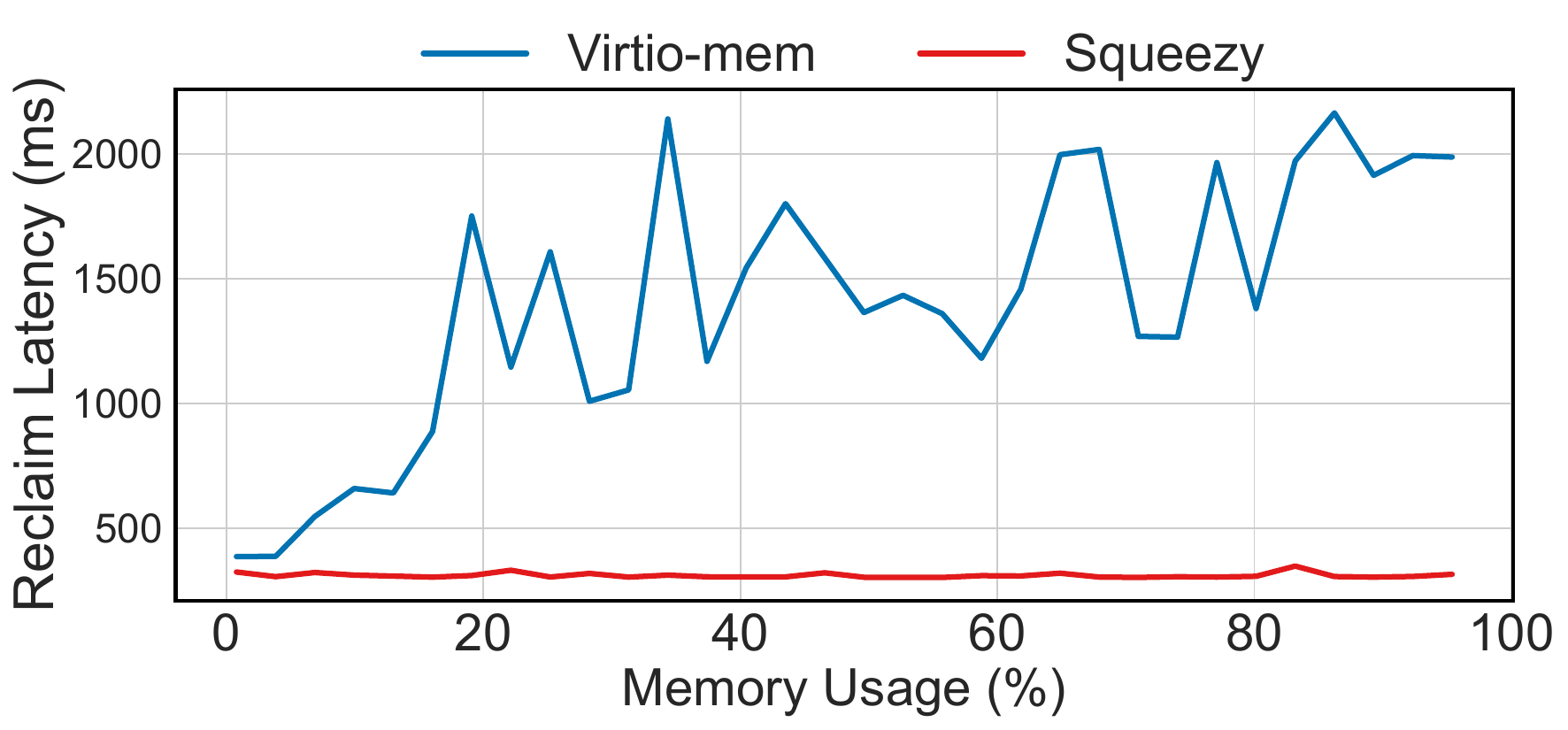} 
   \end{subfigure}
\caption{Reclaiming \shepherd{2 GiB} out of a \shepherd{64 GiB} VM while increasing its memory utilization. Squeezy performance is 
robust and decoupled from the availability of free pages.}
\label{fig:2gPercentUsage} 
\vspace*{-5.0mm}
\end{figure}

\noindent\textbf{Sensitivity to memory utilization.}
We observe that the performance of vanilla virtio-mem depends on the number of occupied pages per reclaimed block.
We conduct a sensitivity study. 
Figure~\ref{fig:2gPercentUsage} shows the latency to unplug 2 GiB from a 64 GiB VM while we increase the utilization of the rest of the memory by increasing the number of running memhog instances. For this set of experiments we remove page zeroing overheads from vanilla virtio-mem as well, to isolate the effect of page migrations.

Vanilla virtio-mem unplug latency exhibits an upward trend along with the memory utilization, as the number of potentially occupied pages and thus migrations per memory block increases. We observe that as soon as 20\% of the VM memory is occupied, page migrations start to significantly slow down unplug operations. This is attributed to the random placement of memhog's pages over multiple memory blocks by the guest OS memory allocator (\S\ref{sec:background:linux}) even when the memory pressure is low. This randomness makes unplug performance also unpredictable, i.e., there are fluctuations in unplug latency as we increase the memory utilization. 

Squeezy instead has robust performance decoupled from the guest load. It always reclaims 2 GiB of memory within $\sim$125 ms by instructing the virtio-mem driver to 
unplug the unused memory blocks of empty Squeezy partitions.

\begin{figure}[t!]
    \begin{subfigure}[b]{0.99\linewidth}
        \centering
        \includegraphics[width=0.99\linewidth]{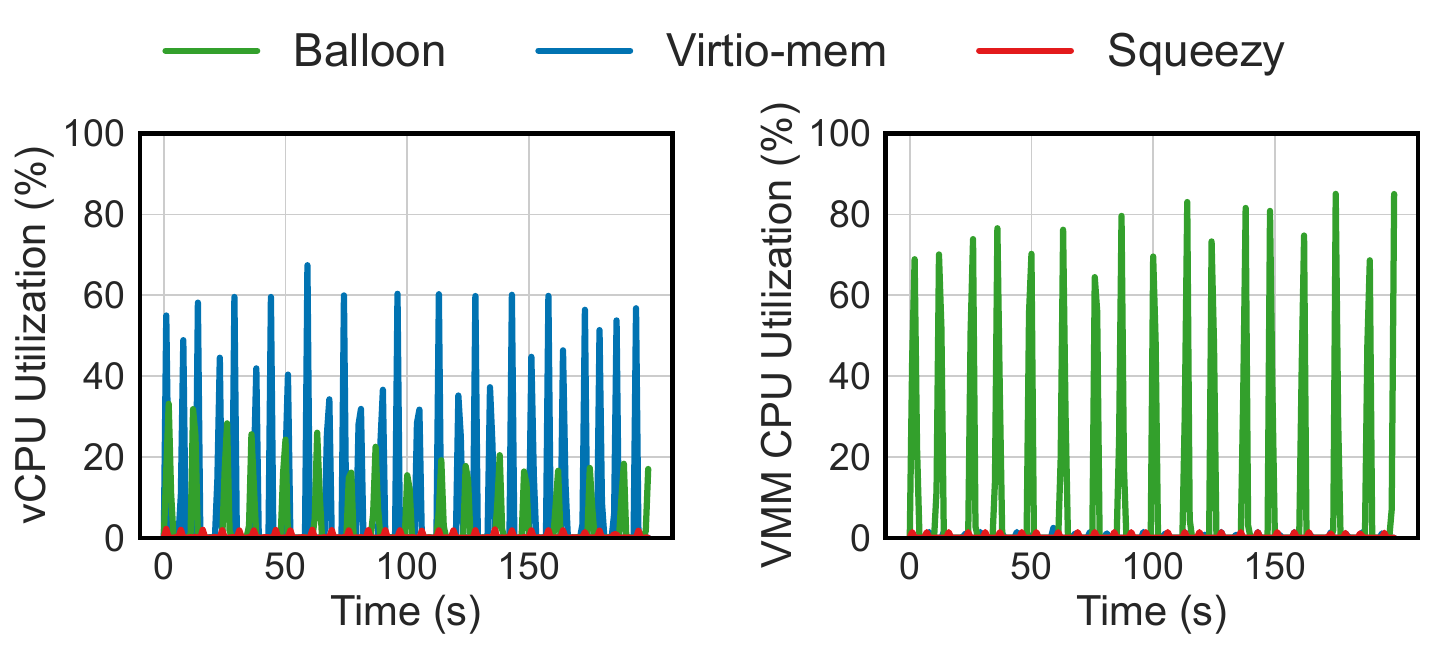} 
    \end{subfigure}
    \caption{
    \revision{
    CPU utilization (\%) of the kernel threads that serve 
    downsizing requests and run in the guest (left) and in the host \shepherd{(hypervisor / VMM, right)} for each method, as we repetitively reclaim 512 MiB of guest memory. Squeezy requires negligible CPU resources to operate.
    }}
    \vspace*{-2.0mm}
    \label{fig:cpuusage}
\end{figure}

\subsubsection{CPU Utilization.}
As discussed in Section~\ref{sec:motiv}, another performance aspect of the mechanisms that reclaim VM memory is the amount of CPU resources that they require to operate. 
For this set of experiments, we add a dedicated vCPU to the VM and pin the guest kernel threads of the different reclamation interfaces, i.e., the thread of the balloon driver and the thread of the virtio-mem driver for vanilla and Squeezy memory hot-unplugging. We also pin the vCPU to a dedicated physical core in the host. Similarly, we pin the host / VMM kernel threads of the interfaces to a separate dedicated physical core. This allows us to isolate and study their CPU activity. Figure~\ref{fig:cpuusage} shows the CPU utilization (\%) of the guest and the host kernel threads as we repeatedly reclaim 512 MiB of VM memory, sleeping between each step, for a 200 second experiment. 

\revision{
The ballooning host kernel thread results in CPU usage spikes while serving the VM exits of the balloon inflation.} Virtio-mem's guest kernel thread heavily uses the vCPU to migrate pages during unplug operations. In a following paragraph, we show how this affects the performance of other processes running in the guest concurrently. 
Squeezy requires negligible CPU resources to reclaim memory and thus minimizes interference.

\subsection{Integration Into a FaaS Runtime}
\label{sec:eval:faas}

In this set of experiments, we study the integration of dynamically resized N:1 VMs  
to a FaaS deployment. We build upon an OpenWhisk-based runtime~\cite{faasmem}, by adding the required functionality to spawn functions as containers inside VMs and resize the VMs on demand based on the incoming load (\S\ref{sec:runtime}). 
We study the FaaS functions of Table~\ref{tab:functions}. We deploy each function inside a dedicated VM, following the N:1 VM model
(\S\ref{sec:motiv:isolation}). 
In our experiments, we always have a single N:1 VM per function type, and we calibrate the max concurrency (N) of the VM to match the maximum 
number of alive instances maintained by the runtime for every trace of requests. N ranges from 9-36. 

The serverless runtime in the host schedules incoming requests to the N:1 VMs, and a dispatcher (\emph{Agent}) within each VM ensures their execution. When there are no available idle function instances (i.e., containers) to serve an incoming request, the Agent creates a new instance (\emph{scale up event}). 
After the request runs to completion, the Agent keeps the function instance cached (\emph{keep-alive}) for a fixed amount of time (2 minutes). 
When new requests arrive, the Agent reuses the cached idle function instances. 
When the keep-alive window expires, the Agent evicts the function instances that were not reused during that window (\emph{scale down event}).
The Agent is also in charge of sharing the CPU resources among the running instances based on their pre-configured CPU share limits (Table~\ref{tab:functions}).

We extend the runtime to support the dynamic up and down-sizing of the VM memory during scale up and down events (Figure~\ref{fig:runtime}). 
We add and reclaim VM memory based 
on the number of instances that are created / evicted and the function memory requirements (Table~\ref{tab:functions}). 

\subsubsection{Resizing FaaS VMs Under Realistic Load.} 
\label{sec:eval:faasazure}
We use the Azure Functions Trace 2021 collection~\cite{harvestserverless} to drive the invocations of every function. We select 4 traces with bursty request patterns and map them randomly to our functions.
\revision{
Specifically, we assign each trace to a function of Table~\ref{tab:functions} and for each trace invocation, we generate a request to the FaaS runtime for the function it has been assigned to.} 
For this experiment, the available memory in the host is abundant and can fit all the instances created by the runtime during the entire trace execution. Thus memory elasticity does not directly affect end-to-end performance,
i.e., there is no requirement to reclaim a VM's idle memory to
enable the scale-up event of another VM.
Our target is to study the performance of dynamically resizing the memory of N:1 VMs when they serve realistic FaaS load.

\noindent\textbf{Scaling up.}
The bursts of requests in the traces trigger scale up events that result in memory hot-plug operations. This can potentially penalize the cold start execution of the new instances due to the added plug delays. To study this penalty, 
we compare the cold start execution of a new instance when we plug its memory (virtio-mem and Squeezy) to \shepherd{the cold start latency of} deploying it on a statically long-running 
over-provisioned N:1 VM (no plugging). 
We measure that the plug operation has negligible overhead for both methods -- it costs 35-45ms for all function sizes.
\shepherd{Guest page fault handling incurs no lookup latency; as discussed in \S~\ref{sec:design:squeezymech} each function is assigned a Squeezy partition Id at creation, allowing direct allocation with 0\% impact.}
However, overall cold-start execution on a dynamically resized VM is 3-35\% slower compared to a static VM. 
This indirect penalty stems from 
the slower accesses to freshly plugged memory; they trigger costly VM exits (nested page faults) that map the touched guest pages to physical resources.
Nevertheless, we consider these cold-start costs of memory elasticity manageable. 
In a later paragraph (\shepherd{\S~\ref{sec:eval:comparison}}), we show that they remain significantly lower compared to booting a 1:1 microVM, which also needs to fault in its memory. 

\begin{figure}[t!] 
 \centering
    \begin{subfigure}[b]{0.64\linewidth}
    \centering
 \includegraphics[width=0.99\linewidth]{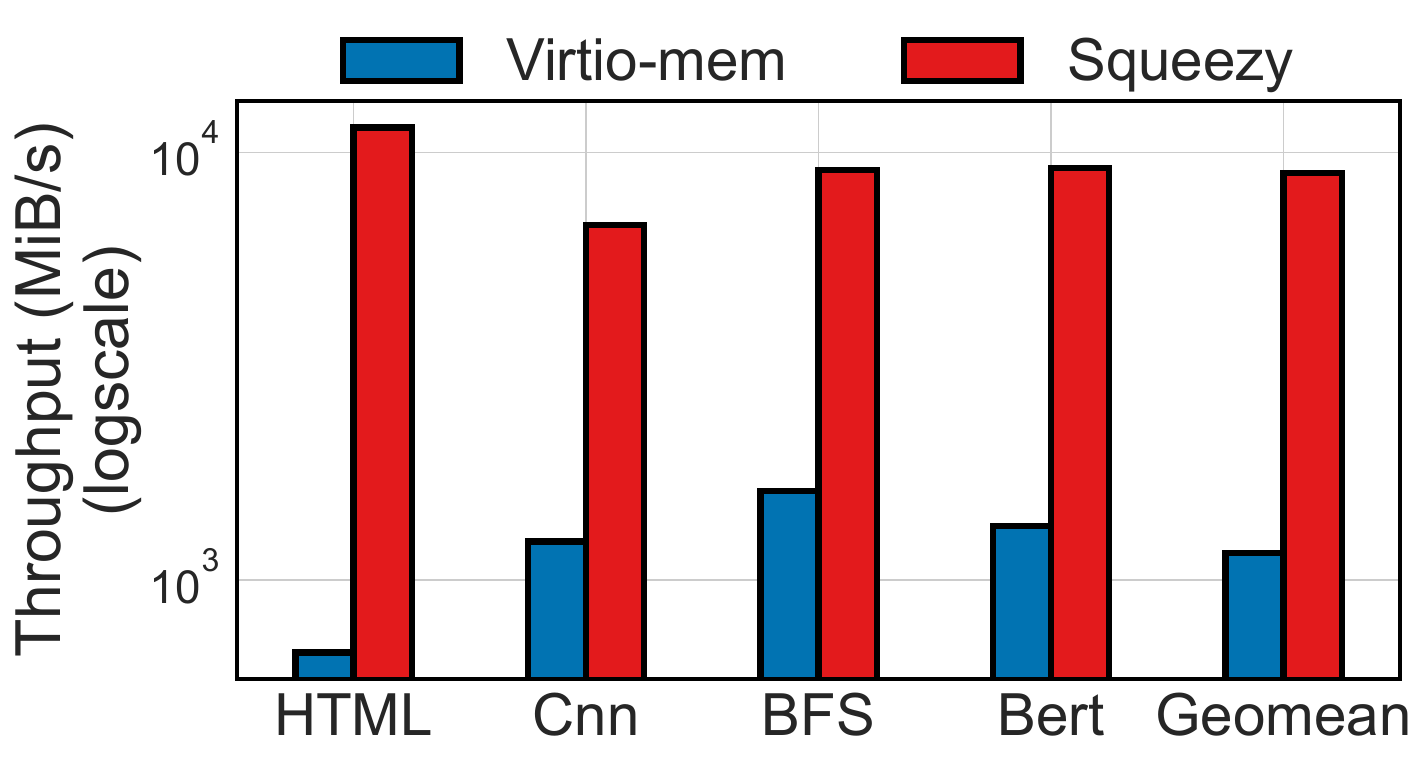} 
    \end{subfigure}
\caption{Memory reclamation throughput (MiB/s) while evicting function instances based on realistic FaaS load.}
\label{fig:faasscaledown} 
\end{figure}

\noindent\textbf{Scaling down.} 
When the load drops in the traces, the FaaS runtime scales down the number of instances, leading to memory hot-unplug events. 
In Figure~\ref{fig:faasscaledown} we report the throughput, in MiB/s, with which vanilla virtio-mem and Squeezy reclaim memory for each function. 
We observe a similar trend to our micro-benchmarking results (\S\ref{sec:eval:ubench:uspeed}), i.e., Squeezy achieves 7x higher reclamation throughput (on average). 

\begin{figure}[t!]
\begin{subfigure}[b]{0.65\linewidth}
    \centering
  \includegraphics[width=0.99\textwidth]{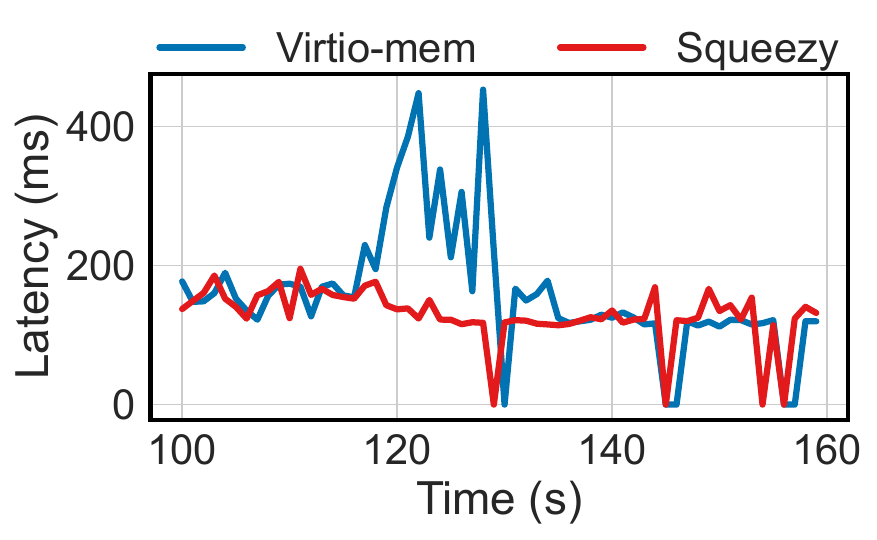} 
\end{subfigure}
\caption{CNN request latency during a scale-down event. Virtio-mem page migrations slow down the running CNN instances by consuming CPU resources. Squeezy instead does not interfere with their execution.}
\label{fig:interference}     
\end{figure}

To further investigate the impact of unplugging events on instances running on the same VM, we co-locate CNN and HTML function instances.
We continue to drive their requests based on the Azure bursty traces. However, the load of the two functions fluctuates at different time intervals and around the $\approx$125s mark of the experiment the FaaS runtime scales down the HTML instances while multiple CNN instances continue to run in the VM, serving requests. Figure~\ref{fig:interference} shows the average end-to-end latency for each second for the CNN requests served, around the minute of the scale-down event. Vanilla virtio-mem slows down requests by more than 2x. The virtio-mem driver's guest kernel thread has been scheduled by the OS to migrate pages using one of the vCPUs that are also assigned to the CNN instances. The CPU utilization caused by the migrations (Figure~\ref{fig:cpuusage}) penalizes their performance.
Squeezy's minimal CPU usage, on the other hand, avoids
any interference with the running instances.

\revision{
\subsubsection{End-to-end Execution When Memory is Limited.}
\label{sec:eval:faase2e}
We now restrict the available host memory for our experiment and synthetically generate load for each function to emulate the spawn and reclaim patterns of a large scale real-world trace (Figure~\ref{fig:motiv:instances})
in our smaller-scale setup (4 functions, single host).
Specifically, we study the end-to-end performance 
when memory is scarce and scale up events \shepherd{have to} 
actively reuse the released memory of 
concurrently scaled-down (evicted) idle instances.
When the available memory in the system is not enough to scale up instances at the rate that the runtime demands, based on the incoming load, the latter attempts to evict as many idle instances 
as necessary and reclaim their memory to proceed.  
For this set of experiments, memory reclamation efficiency can affect the performance at the tail, as scale up events may have to wait for scale down events to finish to secure the necessary memory resources to proceed.
}

\revision{
We normalize performance to the case that we have been studying so far: there is enough free host memory and no reclamation is necessary to secure resources for the hotplug phases of the scale-up events (\shepherd{\emph{Abundant Memory}}). We run the same experiment and we restrict the available memory in the host to $\sim$ 70\% of the maximum memory used in the \shepherd{Abundant Memory} scenario.
We compare Squeezy's dynamic VM resizing to a) vanilla virtio-mem and to b) an 
\shepherd{enhanced} version of virtio-mem, which incorporates optimization techniques from HarvestVM (HarvestVM-opts)~\cite{harvest}.
While the original proposal builds on ballooning, there is no available implementation and the paper targets Windows VMs under Hyper-V. 
We thus isolate and study optimization techniques, proposed by HarvestVM, by applying them to our set-up. 
Specifically we study a) proactive reclamation of memory, i.e., reclaiming 
more memory than necessary for the ongoing scale up events in the system, 
and b) reserving some slack memory (buffering), which can be used by plug events to hide some of the  
delays of slow VM memory reclamation during VM up-sizing.
}

\begin{figure}[h]
\centering
    \begin{subfigure}[b]{0.99\linewidth}
        \centering \includegraphics[width=0.99\linewidth]{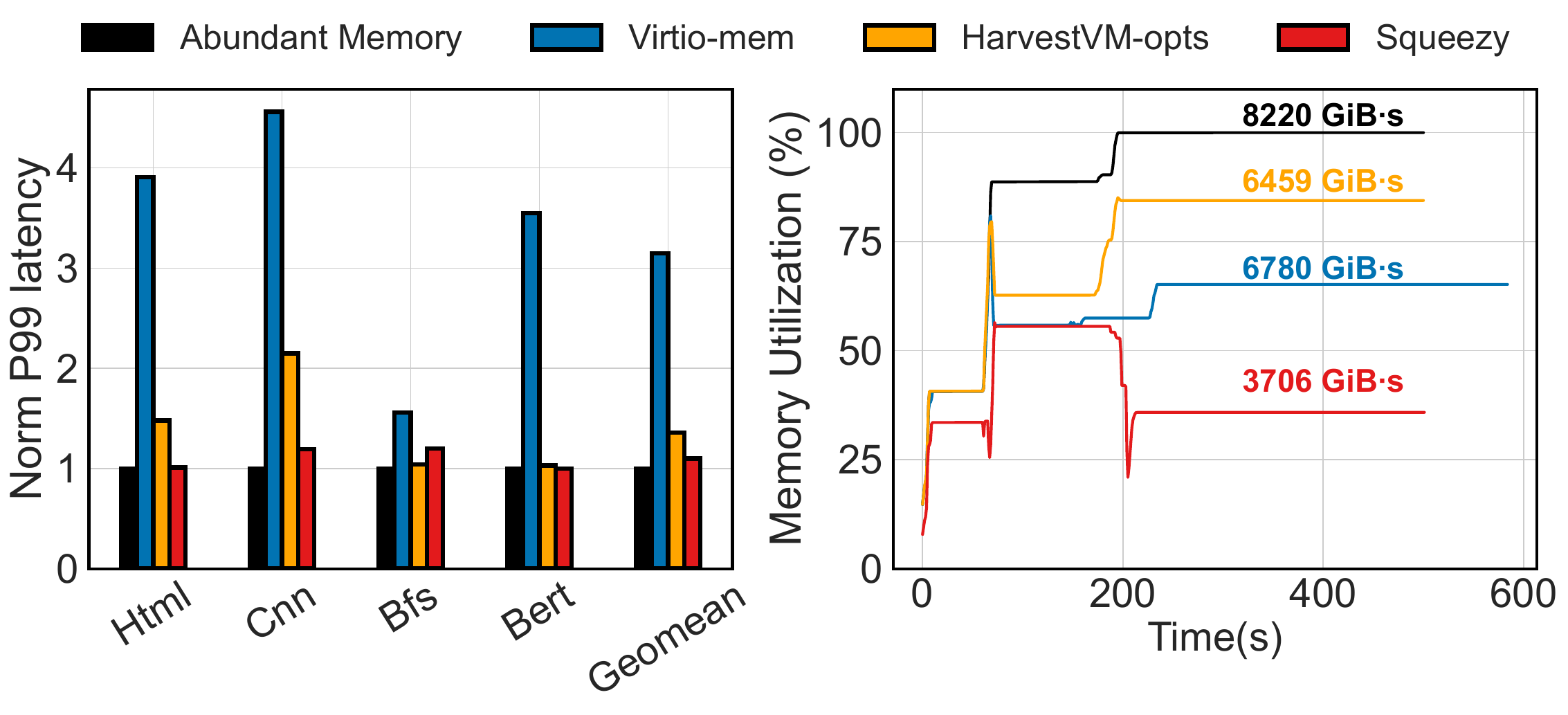} 
    \end{subfigure}
    \vspace*{-2.0mm}
    \caption{\revision{Fast memory reclamation sustains tail latency while saving memory resources.}}
     \label{fig:synth}
\end{figure}

\noindent
\revision{
\textbf{Tail latency.}
Figure~\ref{fig:synth} (left) shows the normalized P99 latency for the entire experiment per function and method. We observe that virtio-mem's slow memory reclamation severely penalizes tail latency (3.15x slower on average),
as scale up events are frequently delayed.
\shepherd{Applying the HarvestVM optimizations improves performance but at the cost of higher memory consumption due to the additional reserved buffers (on the right of Figure~\ref{fig:synth}). It also remains significantly slower compared to the case that pluggable memory is an infinite resource (1.36x on average).}
Squeezy's fast memory reclamation keeps tail latency bounded (1.1.x slower on average), while also minimizing memory utilization, as we describe in the next paragraph. 
}
\shepherd{Moreover, we note that HarvestVM optimizations can negatively impact system robustness due to the more aggressive reclamation of instances when the reserved buffers are full. Depending on the trace request patterns, specific functions might be penalized or favored: for example, we observe that BFS and Bert perform well because they primarily utilize buffered or proactively reclaimed memory from HTML and CNN instances. Conversely, HTML and CNN experience degraded performance due to the premature and aggressive reclamation of their instances, reducing their load capacity and increasing tail-latency. In contrast, Squeezy uses synchronous memory reclamation and does not reserve memory, enhancing robustness and reducing memory waste (on the right of Figure~\ref{fig:synth}). While this incurs minor overhead for BFS (15\%) during scale-ups, it prevents performance degradation for other functions.
}

\noindent
\revision{
\textbf{Memory utilization.}
Finally, we study the memory utilization throughout the duration of the experiment and report the results, in Figure~\ref{fig:synth} (right), normalized to the maximum memory usage reached in the \shepherd{Abundant Memory} scenario, i.e., the maximum footprint when no reclamation is involved. The spikes and drops for the dynamic methods represent the tug-of-war between instance creations (memory allocations / scale up) and evictions (memory reclamations / scale down). 
During scale up events, the HarvestVM-opts method races between allocating reclaimed memory of proactively evicted idling instances ($\approx 90\,\text{s}$) or allocating from the reserved buffer if the reclamation happens at a slower rate than the allocation ($\approx 200\,\text{s}$). With virtio-mem, new instance allocations primarily use reclaimed memory when unplugging occurs quickly  ($\approx 90\,\text{s}$) but when reclamation is slow the method runs into time-outs. 
For example, around the $\approx 200\,\text{s}$ mark, virtio-mem fails to reclaim the necessary memory to scale up the target number of instances in time, delaying the execution of the incoming requests, forcing them to be served by already alive instances, when they eventually become free
later on -- impacting tail latency. 
These reclamation timeouts lead virtio-mem to reclaim less memory than initially targeted, forcing it to use the maximum memory available in the system.
Squeezy reliably reclaims the requested memory just in time, as shown by the drops preceding spikes, and redistributes it to scale up events, keeping memory consumption low.  
Squeezy reduces the overall accounted memory footprint of the functions throughout the duration of the experiment (GiB·s) by 45\% and 42.5\% compared to 
HarvestVM-opts and Virtio-mem respectively.
We note that the higher tail latency of Virtio-mem leads to longer total experiment execution.}

\revision{
\subsection{Comparing the N:1 and 1:1 Models}}
\label{sec:eval:comparison}

\revision{
Finally, we compare the N:1 model, implemented with elastic Squeezy VMs, to the 1:1 model that deploys each function instance to a single microVM. 
We boot the 1:1 microVMs 
with the minimum memory required
to deploy a single function instance (Table~\ref{tab:functions}) 
and assign 1 vCPU to each microVM. 
We compare the two models in terms of: a) cold start delays and b) memory footprint.  
Specifically, we compare booting a 1:1 microVM and creating a new instance to plugging memory on an already running N:1 VM and creating the instance.
}

\begin{figure}[h]
    \begin{subfigure}[b]{0.607\linewidth}
        \centering
        \includegraphics[width=0.99\linewidth]{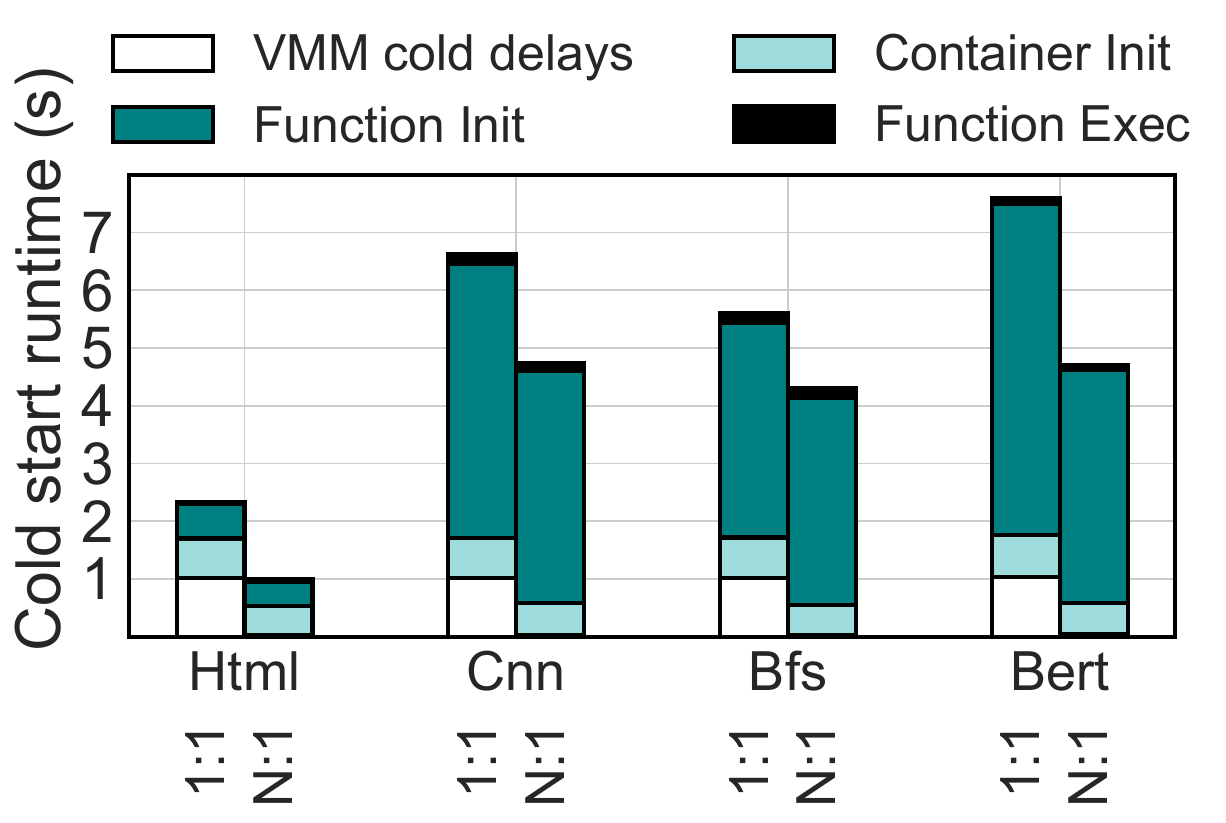} 
        \caption{Cold start latencies.}
        \label{fig:1-1:breakdown}
    \end{subfigure}
    \begin{subfigure}[b]{0.393\linewidth}
        \centering
        \includegraphics[width=0.99\linewidth]{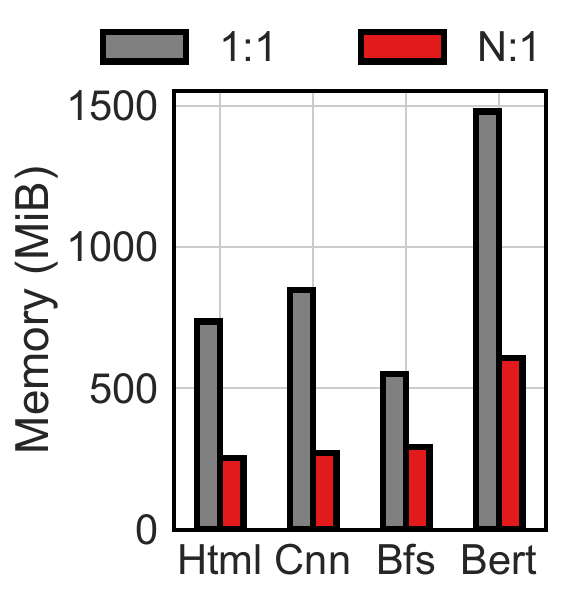} 
         \caption{Memory footprint.}
        \label{fig:1-1:memory}
    \end{subfigure}    
    \caption{\revision{The N:1 and 1:1 model performance trade-offs. Dynamically resized N:1 is implemented with Squeezy.}}
    \label{fig:1-1}
\end{figure}

\revision{
\noindent\textbf{Cold start execution.} Figure~\ref{fig:1-1:breakdown} breaks down cold-start execution to a) VMM cold delays, b) the sandbox initialization (container), and c) the function initialization and execution ~\cite{faasmem}. The VMM overhead corresponds to boot delays for
the 1:1 model and the 
memory plug latency for the N:1.
}

\revision{
We observe that boot overheads slow down 1:1 cold start execution by 20.2\% on average. Deploying a new instance on an already running N:1 VM eliminates this cost and the model's VMM overhead (plug memory) is only 1.19\% on average. We also observe that compared to the 1:1 VM cold run, the container initialization in the N:1 VM is 1.33x faster and the function initialization and execution is 1.25x and 1.08x faster respectively (on average).
The reason for these speed-ups is that the N:1 model shares some state among the running instances, e.g., the container file system and the runtime dependencies are cached and shared via the guest OS page cache.
Thus the initialization of new instances finds some data already in the guest memory and executes faster.
Overall cold start execution is up to 2.35x and on average 1.6x faster when we scale up instances by dynamically resizing a N:1 VM compared to using a new 1:1 microVM per instance. 
}

\revision{
\noindent\textbf{Memory consumption.}
The N:1 model's state sharing across instances is also depicted when we measure the memory footprint of a new instance in the host. Figure~\ref{fig:1-1:memory} shows that instances occupy 2.53x more memory (on average) when they are deployed on a separate 1:1 microVM. This overhead stems from replicating the guest OS state and the function and FaaS runtime dependencies for every instance. Workloads with larger dependencies (e.g., Bert) suffer the most.
}

\section{\revision{Discussion}}
\label{subsec:discussion}
\revision{
\noindent{\textbf{Going beyond FaaS.}}
While Squeezy mainly targets the FaaS use-case, it is also applicable to non-FaaS scenarios. 
The key insight behind Squeezy is that FaaS workloads have predictable memory requirements. 
Each function instance is pre-configured with a maximum amount of memory. 
Cloud-native applications, designed as a mesh of interconnected microservices, also fulfill the above requirement. 
Microservices are commonly deployed as containers on Kubernetes clusters, configured with a maximum amount of memory~\cite{k8s-cgroups}. 
For microservices deployed in VM-sandboxed containers~\cite{kata}, Squeezy could be used to reap the benefits of rapid and lightweight VM memory reclamation, when microservice instances terminate, obviating the need for VM over-provisioning.  
}

\revision{
\noindent{\textbf{Static partitioning.}} 
For longer-running workloads, with less predictable memory requirements, the static partitioning scheme of Squeezy might not be optimal and it would need to be extended, to allow for the plugging and unplugging of variably-sized partitions. 
The trigger for plugging and unplugging would also need to change and be controlled by the application running inside the VM instead. 
}
\revision{We note, however, that for workloads with predictable memory footprints, the static partitioning approach of Squeezy does not imply static memory allocation. 
Squeezy partitions enable the OS to bound, temporally and spatially, the memory allocations of processes, but they do not entail any memory allocations.  
Memory in each partition is allocated on-demand, via nested page faults, in page granularity (4KiB or 2MiB), as the Squeezy-enabled process touches its memory (\S\ref{sec:overview}, \S\ref{sec:design:squeezymech}).}

\revision{\noindent{\textbf{Using Squeezy to optimize keep-alive instances.}} 
We consider for future work integrating the concept of soft-state and soft memory~\cite{midas,softmem} with Squeezy. 
Applications could request Squeezy partitions to use as soft-memory, in order to store application controlled soft-state.
Under memory pressure, the hypervisor could rapidly reclaim soft-memory Squeezy partitions.
Similarly to soft-state, the rapid VM reclamation, enabled by Squeezy, could be used to reclaim unused memory of garbage-collected runtimes, such as Java and Javascript~\cite{frozen}, for VM-sandboxed function instances under the N:1 model. 
Finally, in a similar direction, we consider for future work to extend Squeezy and integrate the concept of temporal segregation of function memory footprints, introduced by FaaSMem~\cite{faasmem}, into the Squeezy partitioning scheme. We will thus be able to extend the Squeezy VM reclamation benefits to function invocations as well as function instance creations and evictions.}

\revision{\noindent{\textbf{Maximum concurrency.}}
As discussed in Section~\ref{sec:implementation}, the concurrency factor $N$ of each N:1 FaaS VM is selected and configured by the runtime. 
Squeezy can accommodate any concurrency factor, without affecting performance.
$N$ effectively acts as an upper bound of the maximum number of function instances that can concurrently be scheduled on the VM. 
Squeezy partitions start initially empty and are only populated via plug events, thus not affecting VM boot performance. 
The effective concurrency factor (i.e., the number of actually populated Squeezy partitions) fluctuates as function instances are created and evicted on the N:1 VM, leading Squeezy to plug and unplug function memory, in fixed sized chunks. 
Recent research~\cite{sesame25} studies the effect of the concurrency factor for the N:1 model and proposes a hybrid approach, that involves VM cloning. Squeezy could be seamlessly used with this approach as well. 
}
\section{Related Work}
\label{sec:related}

\noindent{\textbf{Memory elasticity.}}
To realize memory elasticity on physical machines, Infiniswap~\cite{infiniswap} and FastSwap~\cite{fastswap} rely on memory disaggregation implemented via a swapping-based interface over RDMA fabrics. 
AIFM~\cite{aifm} replaces the kernel-level swapping with fine-grained userspace-controlled swapping to remote memory. 
FluidMem~\cite{fluidmem} targets VMs instead, and improves upon the swap-based approach by utilizing the Linux userfaultfd interface~\cite{userfaultfd}.
All of the aforementioned works achieve memory elasticity by targeting disaggregated RDMA-accessible memory.
By contrast, Squeezy targets FaaS microVMs and, by providing reliable and fast memory reclamation, enables the coupling of memory hot-unplugging with function invocation.

Memory ballooning~\cite{hotplugorballon,resizingballoon,balloonvmware,appballoon,meb,hubballon} has been the state-of-practice interface to implement VM memory elasticity.
The default interface suffers from severe management costs, as it adds and removes memory in the granularity of a page~\cite{resizingballoon, analysis, virtio-mem}. 
A recent work~\cite{mhvm} 
employs a large set of optimizations to work around the method's high latency, e.g., reserving idle memory buffers to fall back to when balloon memory reclamation is late to secure free blocks of memory. We compare Squeezy memory hotunplugging with ballooning in Section~\ref{sec:eval}.

\shepherd{
HyperAlloc~\cite{hyperalloc}, a recent work developed concurrently with Squeezy, presents an efficient mechanism for virtual machine memory elasticity. Compared to our work, HyperAlloc targets large, long-lived IaaS VMs, while Squeezy focuses on short-lived serverless functions. This is reflected in HyperAlloc’s design choices: (i) per-page memory reclamation, while Squeezy matches FaaS per-instance memory reclamation, and (ii) the use of 2 MiB large pages for efficiency, which are rarely used in FaaS. HyperAlloc also replaces the Linux physical memory allocator, while Squeezy integrates with the default Linux buddy allocator.
}

Swap-based approaches~\cite{vswapper,memflex} have also been used to improve VM memory elasticity by utilizing local (hypervisor) resources (memory), usually synergistically with memory ballooning.
Previously, transcendent memory~\cite{tmem,tmem2} also used a swap-based interface to enable seamless VM memory scaling via the Frontswap mechanism~\cite{frontswap}, while others~\cite{utmem} have used transcendent memory to drive VM memory elasticity directly from userspace.

\noindent{\textbf{Serverless scaling.}}
In order to meet the extreme scalability and elasticity demands of the FaaS serverless computing paradigm, various techniques have been proposed.
One line of research focuses on accelerating container sandboxing for serverless functions~\cite{MXFaas,sock,sand,icebreaker,mitosis,trenv,cxlfork}.
LightVM~\cite{lightvm} showcases how VM-based sandboxing can outperform contai\-ner-based sandboxing via lean virtual machines.
Snowflock \cite{snowflock} proposes VM cloning for scaling cloud workloads. 
Recently, microVM snapshotting techniques have been proposed \cite{catalyzer,vhive,snapbpf,iosnaps,faasnap} in order to accelerate function cold-starts in the 1:1 model. Squeezy builds on top of the N:1 model that re-uses VM sandboxes for the instances of the same user to accelerate scale up events. All optimizations accelerating container sandboxing are orthogonal to Squeezy and it could benefit from them.
Recent works investigate combining the N:1 model with snapshots~\cite{sesame25}. 
Squeezy could be used in such hybrid setups to ensure resource elasticity. 

\noindent{\textbf{Resource harvesting.}}
Harvesting of unused resources has been studied both in the context of virtual machines~\cite{harvest,hvm,smartharvest,snape} and serverless computing~\cite{harvestserverless}.
While these works focus on harvesting idle compute resources, i.e., CPUs, others~\cite{mhvm,libra,ofc,deflation,vmdeflation} also support harvesting idle memory resources. 
\revision{More specifically, \shepherd{MHVM}~\cite{mhvm} evaluates harvesting memory with serverless workloads by introducing a number of optimizations including the preservation of large buffers of idle memory, proactive reclamation, and larger reclamation chunks. 
We incorporate pre-reclamation and buffering into our setup and evaluate in Section~\ref{sec:eval:faase2e}.
Squeezy manages to achieve comparable or better tail latency while dynamically up and down-sizing VMs, redistributing their memory, while avoiding the memory overhead of the HarvestVM approach. 
Regarding the reclamation batching, it potentially reduces the VMexit overhead of ballooning (Figure~\ref{fig:ubench:latency}). 
Note though that for Squeezy, the VMExit reclaim overhead is only $\sim$~3ms per 128MiB chunk. 
That said, we consider batching as a future optimization for Squeezy, in order to further reduce the VMexit overheads, when multiple instances need to be reclaimed concurrently.}

\section{Conclusion}
We present Squeezy, a novel memory management framework, that enables the rapid reclamation of hotplugged memory for N:1 FaaS VMs. 
By segregating hot-plugged from normal VM memory, Squeezy bounds the lifetime of allocations on hotplugged memory to the lifetime of the associated function, ensuring that hotplugged memory is readily reclaimable on function exit.
Squeezy is able to reclaim memory of terminated serverless functions an order of magnitude faster than state-of-the-art,  
eliminating the performance impact of memory down-sizing on co-located function instances, as it totally obviates the need for page migrations on the reclaim path.

\begin{acks}
\shepherd{We thank the anonymous reviewers and the paper’s shepherd, Pierre Olivier, for their valuable comments.}
\end{acks}

\newpage

\bibliographystyle{ACM-Reference-Format}
\bibliography{squeezy-bib}
\end{document}